# TWO NEW VARIABLE STARS OBSERVED IN THE FIELD OF THE EXTRASOLAR PLANET HOST STAR WASP-3

M. DAMASSO[1], A. CARBOGNANI[1], P. CALCIDESE[1], P. GIACOBBE[2], A. BERNAGOZZI[1], E. BERTOLINI[1],
M. G. LATTANZI[3], R. SMART[3], A. SOZZETTI[3]

1) Astronomical Observatory of the Autonomous Region of the Aosta Valley, Fraz. Lignan 39,
11020 Nus (AO) – Italy (www.oavda.it). *E-mail of the corresponding author*: m.damasso@gmail.com
2) Dept. of Physics - University of Turin, Italy
3) INAF - Astronomical Observatory of Turin, Pino Torinese, Italy

**Abstract:** We report the discovery of two new short-period variable stars in the Lyra constellation, GSC2.3 N208000326 and GSC2.3 N20B000251, observed at the Astronomical Observatory of Autonomous Region of the Aosta Valley. Photometric measurements collected during several days are presented and discussed. One star appears to be a δ Scuti pulsating star (P=0.07848018±0.00000006 days; pulsation amplitudes ΔV~0.055 mag and ΔR~0.045; $(V-R)_{average}$=0.378±0.009, probable spectral type F2). The identity of the second star (P=0.402714±0.000008 days) resulted more difficult to be understood. We propose that this object should be classified as an eclipsing binary system where ~0.065 mag and ~0.055 are, respectively, the depths of the primary and secondary minimum in the light curve, as measured with a non standard R filter.

## 1  Introduction

Two new variable stars were observed at the Astronomical Observatory of the Autonomous Region of the Aosta Valley (OAVdA; 45.7895° N, 7.47833° E). The OAVdA is located at 1675 m above the sea level in the Italian Alps, close to the border with France and Switzerland. The discovery occurred during the 'phase A' of a photometric study aimed at establishing the potential of the OAVdA site to host a long-term photometric transit search for small-size ($R<R_{Neptune}$) extrasolar planets **(Damasso et al., 2009)**. As a part of this feasibility study, during the period May-August 2009 and for a total of 19 nights we collected in- and out-of-transit light curves of the star WASP-3 in the Lyra constellation, known to host an extrasolar planet **(Pollacco et al. 2007)**. These new variable stars represent an interesting by-product of our observations of the WASP-3 field.

## 2  Instrumentation and methodology

All observations were carried out with a 250 mm f/3.8 Maksutov reflector telescope mounting a CCD camera Moravian G2-3200ME with a camera sensor area of 2184 x 1472 square pixels (pixel linear dimension 6.8 μm) and a quantum efficiency of ~87% at wavelength of 610 nm. This configuration results in a field of view of 52.10' x 35.11' and a plate scale of 1.43 arcsec/pixel. All data were taken with an R filter (Astronomik) centered at the wavelenght of 610 nm.

In order to collect standard photometric data for the two new variable stars, follow-up observations were conducted using:

- the OAVdA largest optical telescope, an 810 mm f/7.9 Ritchey-Chrétien reflector equipped with a Finger Lakes Instrumentation Pro Line PL 3041-BB back illuminated CCD camera with standard BVRI filters. The camera sensor area is 2048 x 2048 square pixels (pixel linear dimension 15 μm), resulting in a field of view of 16.3' x 16.3', with a plate scale of 1 arcsec/pixel in binning 2x2.
- a 400 mm f/7.64 Ritchey-Chrétien reflector telescope, mounting a front illuminated CCD camera Finger Lakes Instrumentation Pro Line PL1001E with sensor KAF1001E and standard BVRI filters. The camera sensor area is 1024 x 1024 square pixels (pixel linear dimension 24 μm). The field of view is 26.4' x 26.4' with a plate scale of 1.55 arcsec/pixel.

To calibrate the V and R magnitudes we used the method described by **Dymock and Miles, 2009** and magnitudes reported in the CMC14 (*r'* magnitude) and 2MASS catalogues (*J* and K magnitudes).

The data were reduced and elaborated with the automatic pipeline TEEPEE (Transiting ExoplanEt PipElinE) that we have developed and tested as a part of our photometric study focused on transiting extrasolar planets. TEEPEE performs image calibration (dark and flat-fielding corrections, rotation and translation to the same pixel grid using a reference image), airmass corrections (using an extinction coefficient averaged over typically dozens of stars) and ensemble differential aperture photometry, using typically up to 100 reference objects (depending on the target field) with apparent magnitudes close to the target magnitude.

To confirm that these variables are really a new discovery, we searched through several sources: the General Catalogue of Variable Stars (GCVS, http://www.sai.msu.su/groups/cluster/gcvs/gcvs/), the New Catalogue of Suspected Variable Stars (http://www.sai.msu.su/groups/cluster/gcvs/gcvs/nsv/) and in VizieR database (http://vizier.u-strasbg.fr/). Moreover, we also searched through the lists of new periodic variable stars reported by the team of the SuperWASP project (**Street et al., 2006; Norton et al., 2008**).

We adopted **Sterken and Jaschek (1996)**, **Breger (2000)** and **Percy (2007)** as the main references to suggest a classification for the new variable stars we observed.

## 3   Results and discussion on the individual objects

In Table 1 we summarize the positions, variability periods and amplitudes for both stars, and in Table 2 we list the CMC14 stars we used for calibrating the V and R magnitudes. The finding charts showing these stars are presented in Fig. 2 (1$^{st}$ variable) and Fig. 7 (2$^{nd}$ variable).

### 3.1   Star #1 - GSC2.3 N208000326

In Fig. 1 we reproduce a finding chart for this object. This star is not reported as a variable star in any of the catalogues we looked up. Figure 3 shows the phase-folded light curve for this star (normalized flux) obtained using a pulsation period of 0.07848018±0.00000006 days, estimated with the Lomb-Scargle algorithm implemented in the Starlink-PERIOD package **(Dhillon et al., 2001)**. The data presented here refer to 5 observing nights in the period second half of July to second half of September 2009 and were collected with the 250 mm telescope and R non standard filter. The error bars correspond to 1 σ. In Fig. 4 the R and V calibrated light curves are showed, obtained with the 810 mm telescope during the follow-up measurements of the night 28-29 September 2009. The plots show that the pulsation amplitudes are ~0.055 mag in V and ~0.045 mag in R. Figure 5 shows the variation of the V-R index recorded during the same night, with an average value (V-R)$_{average}$=0.378±0.009. The temporal variation appears to be in phase with V and R, strengthening the hypotesis that we are in front of a pulsating star. To correct this value for the interstellar reddening, we used the galactic dust reddening and extinction maps available on the Web at http://irsa.ipac.caltech.edu/applications/DUST/ (the NASA/ IPAC Infrared Science Archive, or IRSA) which are derived from data and techniques described in **Schlegel, Finkbeiner & Davis (1998)**. At the star location (galactic coordinates *l*=63.5919° and *b*=+18.4108°) the V and R corrections are respectively $A_V$=0.205 and $A_R$=0.174 magnitudes (they represent integrated value along the line of sight) corresponding to a V-R color excess E(V-R)=0.031. This leads to the de-reddened estimate for the index color $(V-R)_{average}$=0.347±0.009, which is typical of a F2 type main sequence star, as reported in **Zombek (1990)**.
Considering the shape of the light curve, the values found for the pulsation period, magnitude variations and V-R index, we suggest that this star can be classified as a δ Scuti variable.

We can then estimate the mean absolute visual magnitude of the star using the Period-Luminosity relation determined by **Poretti et al. (2008),** which does not depend on the knowledge of the star metallicity:

$M_V$ = -3.65(±0.07)log (P) - 1.83(±0.08)

where *P* is given in days. We get $M_V$ = +2.20±0.11.
The mean apparent V magnitude (V=12.625±0.005 from our measurements) can be corrected for the interstellar extinction using two different models of the galactic dust distribution, which lead to two independent estimates for the star distance. Using the analytic expression in **Arenou et al. (1992)**, we assume for the interstellar extintion $A_V$ = 0.111 mag (integrated along the line of sight) as estimated for a star with galactic coordinates *l* = 63.5919° and *b* = +18.4108°. The maximum distance of the star results ~1153±26 pc (~3761 l.y.).
Using the IRSA galactic dust reddening and extinction maps we found $A_V$ = 0.205 mag for the integrated galactic extinction along the line of sight at the star location. This leads to a second, independent maximum star distance of ~1104±25 pc (~3600 l.y.).
Taking the average value of the two indpendent results, we then can assume d=1128.5±36 pc as the better estimate for the maximum distance of the star.

### 3.2    Star #2 - GSC2.3 N20B000251

The finding chart containing the second variable star is showed in Figure 6. We did not find in the literature any indication about the variability of this object. This target was observed on the 250 mm telescope (and the Astronomik non standard R filter) for 9 nights in the period June-July 2009. After the first inspection of the star light curve for the whole observing period, the object showed what appeared a pulsation period of ~0.2 days (obtained applying the Lomb-Scargle algorithm in the PERIOD package) and a magnitude amplitude of ~0.03 mag. These results in principle could indicate that the star is another δ Scuti. We then measured V and R magnitudes on 14th October 2009 with the 400 mm telescope obtaining an average V-R index $(V-R)_{average}$=0.55±0.01, corresponding to a fraction of the estimated period (~2 hr). The variations of the V-R index observed during that night is showed in Fig. 8. Then we corrected this value for the effects of the galactic dust using the IRSA maps (galactic coordinates of the target: *l*= 64.4724° and *b*=+18.5112°), and we assumed a color excess E(V-R)=0.03±0.01. If the target were an isolated main sequence star, our best estimate of the V-R index is indicative of a G-type star **(Zombek, 1990)**. As δ Scuti are A-F spectral type stars, we were forced to reconsider our hypothesis on the real nature of this variable.

We reorganized the data using P=0.402714±0.000008 days as the variability period, which is double of the value we initially adopted. In Fig. 9 we show the corresponding normalized phase-folded light curve. The error bars correspond to 1 σ. From the plot it results clear that, using the new period, the object shows two distinct minima which differ only for ~0.01 mag, while the maxima differs slightly in shape and havesimilar values. This clearly reveals that the object is a binary system. To note that the light curve did not show evident variations in its shape during the two-months period of our observations.

Only on the base of our photometric data some doubts necessairly remain regarding the precise identity of the system, and we considered here two plausible alternatives.

The object could be a short-period non eclipsing binary system showing the properties typical of an ellipsoidal variable (ELL variable star according the GCVS classification). In this case, the small changes in the luminosty are due to the so called gravity darkening induced by gravitational tidal distortion induced by the presence of a much less luminous compact object as companion, which does not contribute to the observed radiative flux **(e.g. Beech, 1985)**. Besides, the system has a measured orbital period which falls in the range showed by cataclysmic variables, close to the upper limit **(e.g. Percy, 2007)**.

The second possibility is that the object is a short-period, grazing eclipsing binary system where the primary and secondary stars could be assumed to be main sequence stars of different spectral type.

Our opinion is that the latter hypothesis is much more reliable and it is based on the following clues. In Table 3 we report the V, J, H and K magnitudes for this object as given by the GSC2.3 and 2MASS catalogues. The V-K color index for this target is higher than expected for a G-type star, while it is typical of a late K-type star **(Zombek, 1990)**. If we are observing an elongated star orbiting a compact object, the information we get from the V-R and V-K color indexes, concerning the spectral type of the secondary star, is difficult to explain in terms of just a difference in temperature between the hotter emisphere that points towards the primary star (which is supposed to be colder) and the opposite. In this case, we should observe a redder object (K-type) when the secondary star shows us the colder emisphere, and it should appear as a G-type when it is in inferior conjunction. We did not find in literature examples of secondary stars in close binary systems which show ellipsoidal variations associated with their filled Roche lobe and which are characterized by an excursion in their spectral type as wide as the one observed for our target.

We believe that the simplest and probably true scenario is assuming the system composed of two normal stars, one that is cooler and brighter in red/near-infrared bands -then determining the observed high value for the V-K index- while the hotter companion is brigther at shorter wavelenghts, producing a V-R index shifted towards an upper spectral class.

As a further check to support our interpretation, we also looked up the HEASARC catalogue (http://heasarc.gsfc.nasa.gov/, a very extended Web-based archive for gamma-ray, X-ray, and extreme ultraviolet observations of cosmic sources) for a possible X-ray counterpart of the target, giving hints of the presence in the binary system of a compact object and of an exchange of matter with the secondary star, the donor companion. We did not find any source at the variable star location which is associated to high-energy emissions. This evidence is in favour of our proposed scenario.

Because we do not have any spectroscopic information on the object, we can not guess anything more about the real nature of the stellar members of the system. Therefore we will not go further in our discussion.

# 4 Conclusions

We reported the discovery of two new short-period variable stars in the Lyra constellation. They represent interesting by-products of a campaing finalized to the observation of extrasolar planets with the photometric transit method.

The objects were monitored for several days in order to better characterize them photometrically and to guess the properties of their variability. According to our results, we suggest that one star can be classified as a δ Scuti variable with asymmetric light curve (for which we estimated the maximum heliocentric distance: d=1128.5±36 pc). The second object should be classified as a detached eclipsing binary system but spectroscopic data are necessary to definitively confirm this hypothesis.

# 5 Acknowledgments

This work has been possible thanks to the contributions of the European Union, the Autonomous Region of the Aosta Valley and the Italian Department for Work, Health and Pensions. We used the following resources: the VizieR catalogue access tool (CDS, Strasbourg, France); the General Catalogue of Variable Stars (GCVS), mantained by the Sternberg Astronomical Institute (Moscow, Russia); the Northern Sky Variability Survey, operated by the University of California for the National Nuclear Security Administration of the US Department of Energy; the NASA/ IPAC Infrared Science Archive, which is operated by the Jet Propulsion Laboratory, California Institute of Technology, under contract with the National Aeronautics and Space Administration; the High Energy Astrophysics Science Archive Research Center (HEASARC), provided by NASA's Goddard Space Flight Center.

We thank Ennio Poretti (INAF-Osservatorio Astronomico di Brera) for the help in understanding the nature of the second object discussed in this paper.

*Table 1. Summary of the main data for the two new variable stars.*

| Star number | Catalogue IDs | Coordinates (J2000, GSC2.3 Catalogue) | Period (days) | Magnitude at max. brightness | Magnitude at min. brightness | Suggested variability type (according the GCVS) |
|---|---|---|---|---|---|---|
| 1 | GSC2.3 N208000326 (also CMC14 183429.5+350424 ; 2MASS 18342951+3504242) | ($\alpha$) 18:34:29.51 ($\delta$) +35:04:24.3 | 0.07848018 ± 0.00000006 | 12.57±0.005 (Johnson-Cousins V) 12.20±0.005 (Johnson-Cousins R) | 12.68±0.005 (Johnson-Cousins V) 12.29±0.005 (Johnson-Cousins R) | DSCT with asymmetric light curve |
| 2 | GSC2.3 N20B000251 (also CMC14 183528.6+355325 ; 2MASS 18352859+3553255) | ($\alpha$) 18:35:28.62 ($\delta$) +35:53:25.3 | 0.402714 ± 0.000008 | 13.69±0.01 (Johnson-Cousins V) 13.15±0.01 (Johnson-Cousins R) | Not available | E |

*Table 2. Data of the field stars used to calibrate the V and R magnitudes of the new variable stars, according to the procedure described by **Dymock and Miles, 2009**.*

| Variable star number | Catalogue ID for the field star | Coordinates (J2000, CMC14 Catalogue) | r' mag (CMC14) | J-K index (CMC14) |
|---|---|---|---|---|
| 1 | CMC14 183449.4+351240 | ($\alpha$) 18:34:49.477 ($\delta$) +35:12:40.75 | 11.69 | 0.631 |
| 1 | CMC14 183428.4+350219 | ($\alpha$) 18:34:28.435 ($\delta$) +35:02:19.81 | 12.434 | 0.317 |
| 1 | CMC14 183402.0+350401 | ($\alpha$) 18:34:02.082 ($\delta$) +35:04:01.54 | 12.621 | 0.367 |
| 1 | CMC14 183451.4+350941 | ($\alpha$) 18:34:51.494 ($\delta$) +35:09:41.13 | 12.705 | 0.556 |
| 2 | CMC14 183527.7+354601 | ($\alpha$) 18:35:27.794 ($\delta$) +35:46:01.46 | 12.277 | 0.68 |
| 2 | CMC14 183550.1+354546 | ($\alpha$) 18:35:50.114 ($\delta$) +35:45:46.56 | 12.704 | 0.331 |
| 2 | CMC14 183558.2+354717 | ($\alpha$) 18:35:58.238 ($\delta$) +35:47:17.85 | 13.585 | 0.687 |
| 2 | CMC14 183532.9+354927 | ($\alpha$) 18:35:32.926 ($\delta$) +35:49:27.39 | 13.35 | 0.394 |

*Table 3. Photometric data of the second variable star as given by the GSC2.3 (V magnitude) and 2MASS (J, H, and K magnitudes) catalogues.*

| **V[a]** | **J** | **H** | **K** | **V-K[a]** |
|---|---|---|---|---|
| 13.57±0.36 | 11.870±0.021 | 11.358±0.029 | 11.274±0.025 | 2.296±0.360 |

[a] According our calibrated magnitudes, limited to ~2 hr of observations, we can assume V=13.73±0.01 as the better estimate for the average V magnitude. This is in accordance with the value reported in the GSC2.3 catalogue.
If we assume our observed value for V and a color excess E(V-K)=0.158 as estimated using the IRSA galactic dust maps, the V-K color index results to be 2.298±0.027, in agreement with the value reported in this Table.

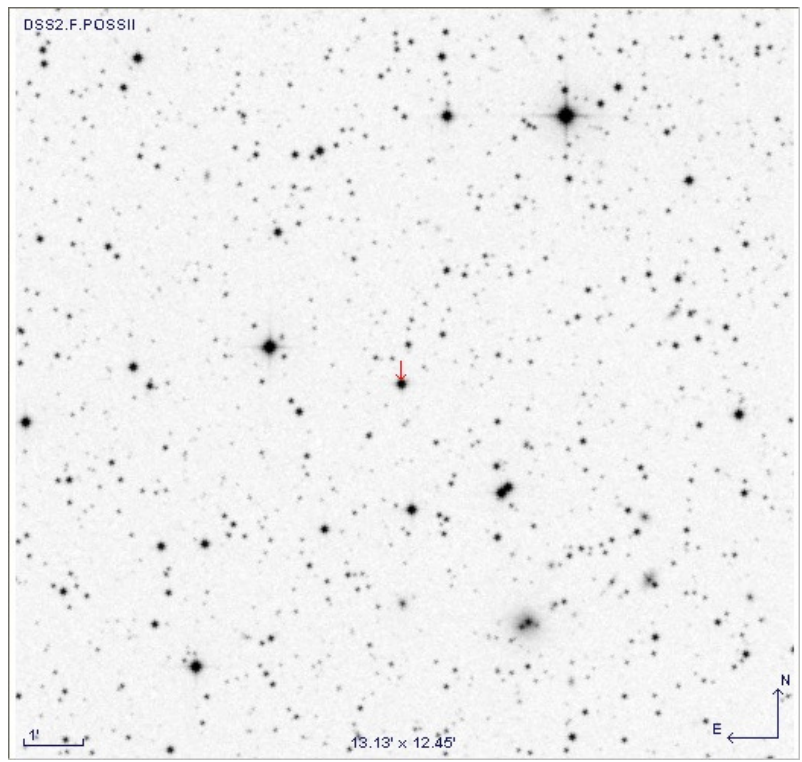

*Fig. 1. Image from the ESO Digitized Sky Survey database showing the star field of the variable star GSC2.3 N208000326, which is indicated by a red arrow.*

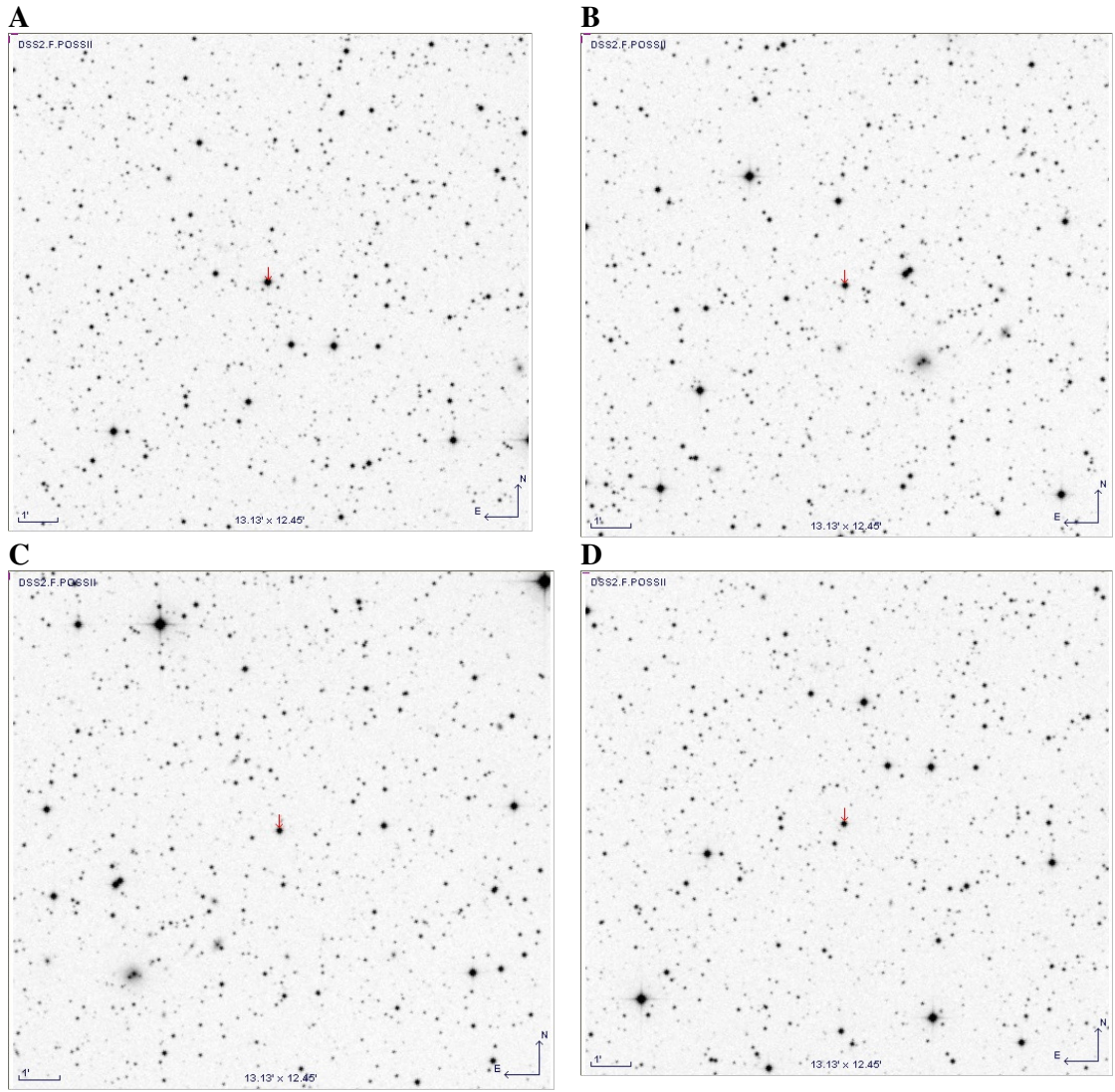

*Fig. 2. Finding charts showing the field stars used to calibrate the V and R magnitudes of the first variable star (indicated by a red arrow at the center of each image), as listed in Table 2. **A:** CMC14 183449.4+351240; **B:** CMC14 183428.4+350219; **C:** CMC14 183402.0+350401; **D:** CMC14 183451.4+350941.*

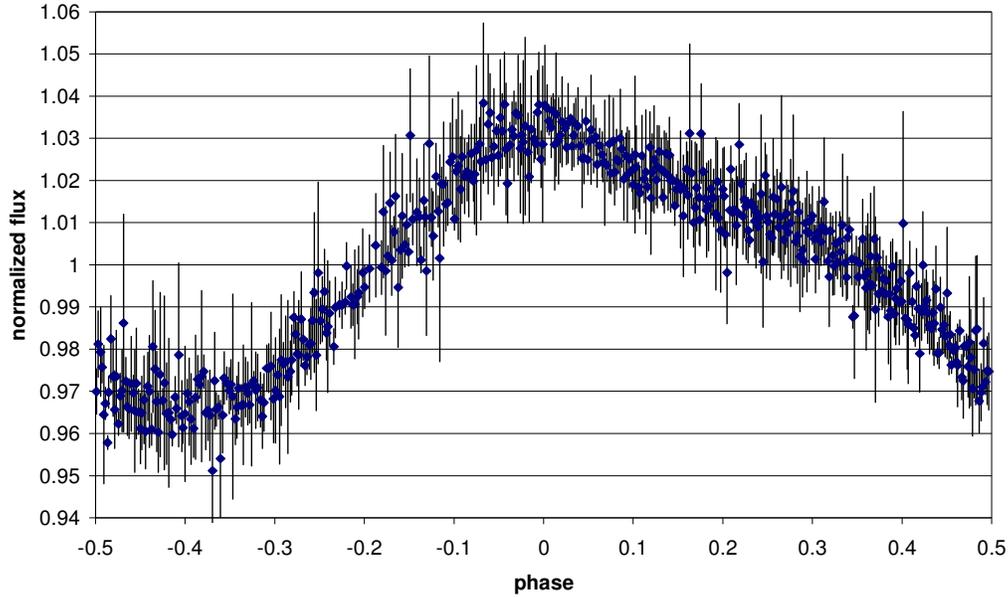

*Fig. 3. Normalized phase folded light curve of the variable star GSC2.3 N208000326, obtained using P=0.07848018 ± 0.00000006 days as the better estimate for the fundamental pulsation period (phase=0 corresponds to ephemeris HJD=2455103.3939+P\*E). The asymmetric shape of the light curve is clearly visible.*

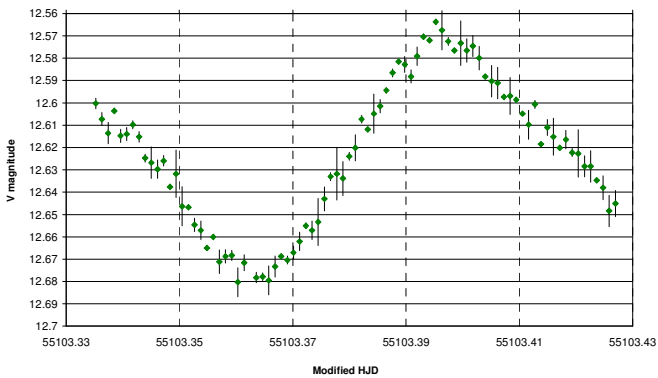
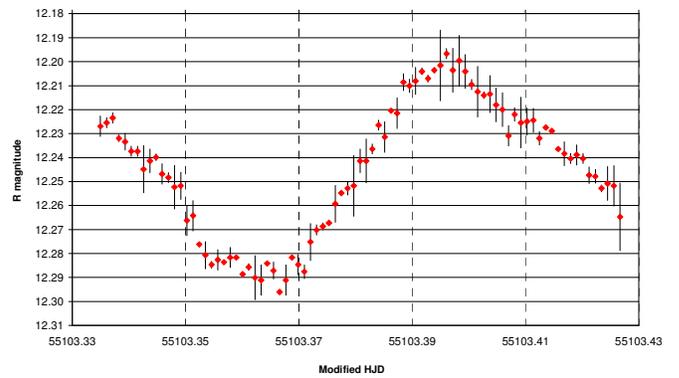

*Fig. 4. V and R calibrated light curves of the pulsating star GSC2.3 N208000326 obtained during the night 28-29 September 2009, when the object was monitored with the 810 mm telescope.*

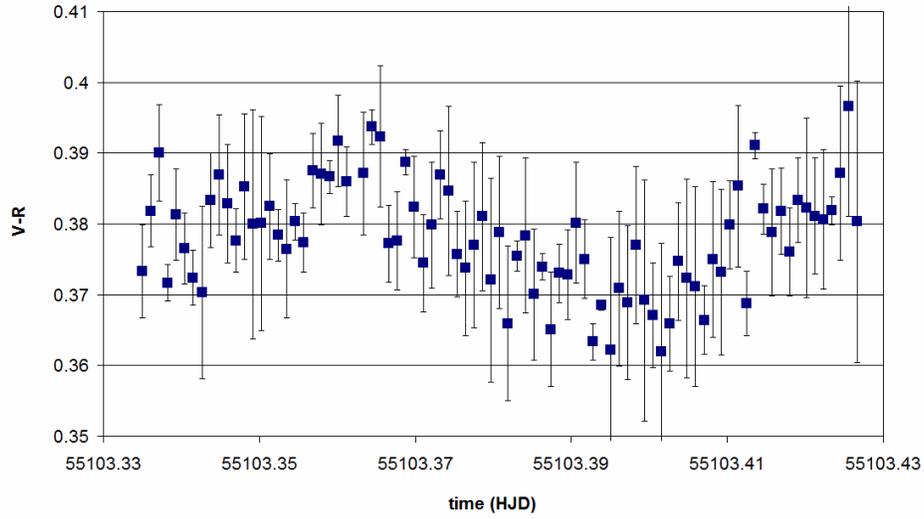

*Fig. 5. Variation of the color index V-R observed for the star GSC2.3 N208000326 during the night 28-29 September 2009.*

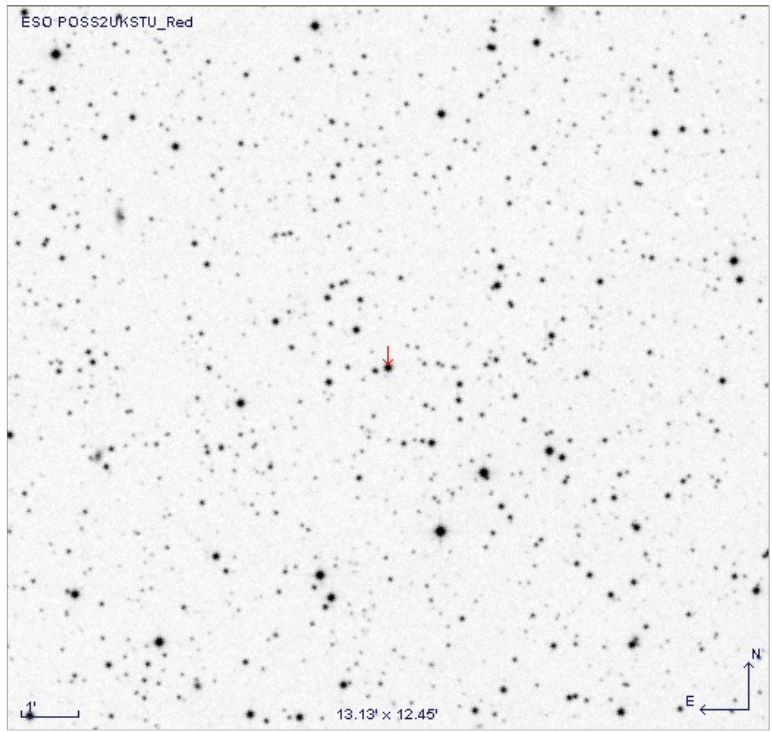

*Fig. 6. Image from the ESO Digitized Sky Survey database showing the star field of the variable star GSC2.3 N20B000251, which is indicated by a red arrow.*

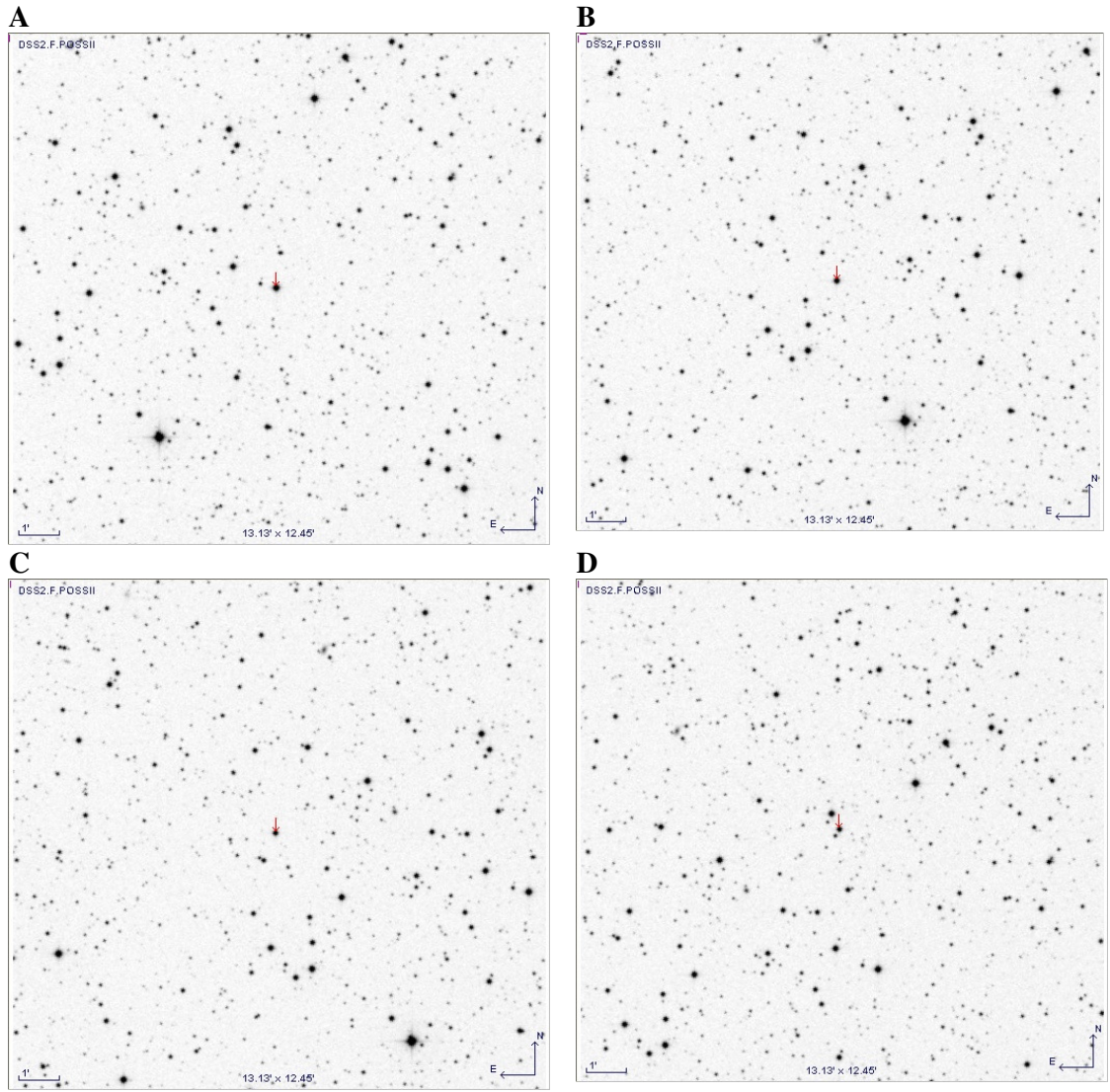

*Fig. 7. Finding charts showing the field stars used to calibrate the V and R magnitudes of the second variable star (indicated by a red arrow at the center of each image), as listed in Table 2. **A:** CMC14 183527.7+354601; **B:** CMC14 183550.1+354546; **C:** CMC14 183558.2+354717; **D:** CMC14 183532.9+354927.*

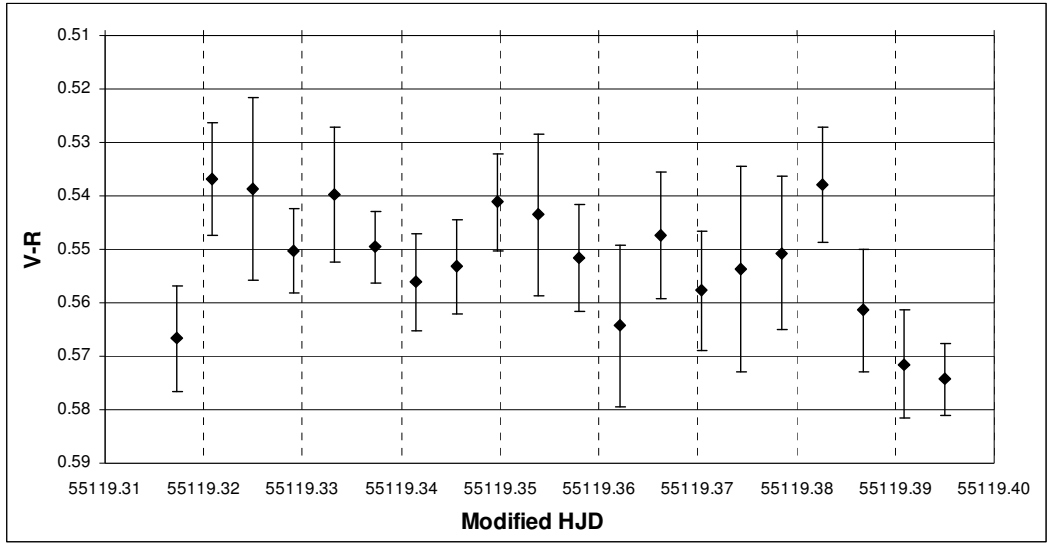

*Fig. 8. Variation of the color index V-R observed for the star GSC2.3 N2B000251 during the night 14-15 October 2009.*

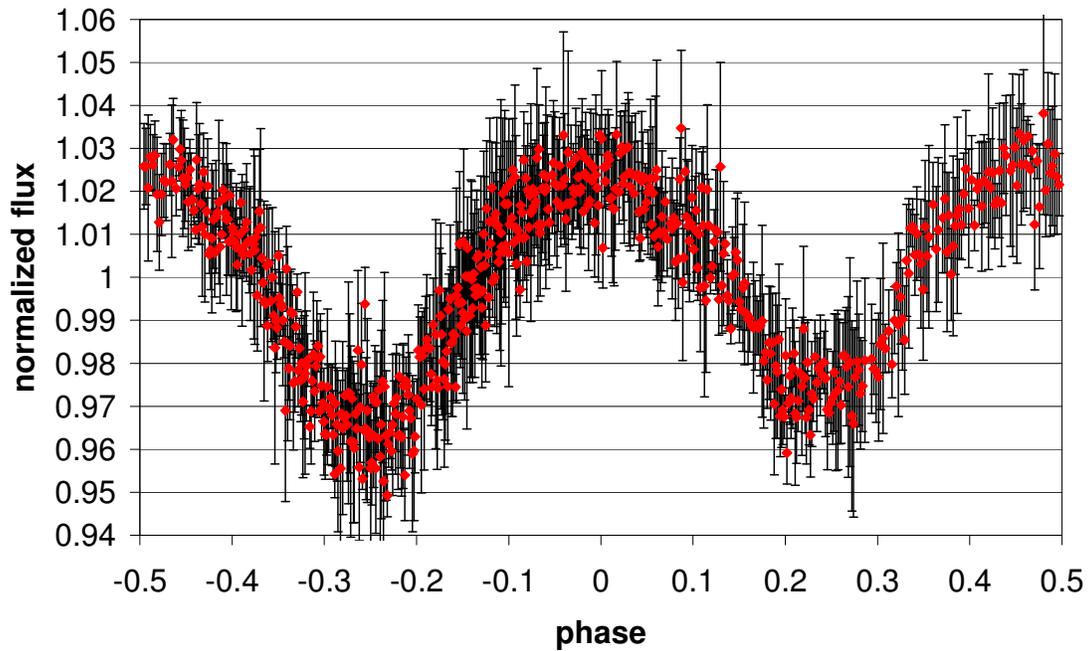

*Fig. 9. Normalized phase-folded light curve of the variable star GSC2.3 N20B000251, obtained using P= 0.402714 ± 0.000008 days as the better estimate for the orbital period of the target (phase=0 corresponds to ephemeris HJD=2454994.501335+P*E).*